\documentclass[11pt, a4paper]{book}

\usepackage[palatino]{anuthesis}

\usepackage[final]{prelim2e}  % use this version to disable the footer

\usepackage[square]{natbib}

\usepackage[T1]{fontenc}
\usepackage[scaled]{PTSans}          % use Paratype Sans for sans-serif
\usepackage[scaled]{DejaVuSansMono}  % use DejaVu Sans Mono for monospace

\usepackage[stretch=10,shrink=10]{microtype}
\usepackage{amsmath}
\usepackage{amssymb}

\usepackage{graphicx}
\usepackage[table]{xcolor}
\usepackage{float}

\usepackage{caption}
\usepackage{subcaption}
\usepackage{booktabs}
\usepackage{relsize}
\usepackage{multirow}
\usepackage{array}
\usepackage{listings}

\usepackage{enumitem}

\definecolor{todocolor}{rgb}{0.8,0.8,1.0}
\usepackage[color=todocolor,colorinlistoftodos]{todonotes}

\usepackage[binary-units=true]{siunitx}

\usepackage{xspace}
\usepackage{afterpage}
\usepackage[normalem]{ulem}
\usepackage{makeidx}
\usepackage{url}
\usepackage{pdflscape}
\usepackage[
  hyperindex=true,
  bookmarks=true,
  pdftitle={YOUR TITLE HERE},
  pdfauthor={YOUR NAME HERE},
  colorlinks=false,
  plainpages=false,
  pdfborder={0 0 0},
  pdfpagelabels,
  hyperfootnotes=false]{hyperref}
\usepackage[all]{hypcap}  % fix links to figures to point to top not bottom

\usepackage[hyperpageref]{backref}
\renewcommand*{\backref}[1]{}
\renewcommand*{\backrefalt}[4]{
  \ifcase #1 %
  \or
    (cited on page #2)%
  \else
    (cited on pages #2)%
  \fi
}

\usepackage[nameinlink,capitalise]{cleveref}

\newcommand{\doi}[1]{\href{http://dx.doi.org/#1}{\nolinkurl{doi:#1}}}

\lstdefinestyle{default}{
  numbers=left,
  numberstyle=\tiny,
  stepnumber=1,
  numbersep=2em,
  language=java,                         % the language
  basicstyle=\footnotesize\ttfamily,     % the basic font family to use
  commentstyle=\itshape,                 % the font for comments
  stringstyle=\ttfamily,
}
\crefname{line}{line}{lines}  % cleveref wants to capitalise "Line" mid-sentence

\definecolor{tableheadcolor}{rgb}{0.8,0.8,1.0}
\definecolor{tablealtcolor}{rgb}{0.9,0.9,0.95}

\newcolumntype{B}{>{\global\let\currentrowstyle\relax}}
\newcolumntype{^}{>{\currentrowstyle}}
\newcommand{\rowstyle}[1]{\gdef\currentrowstyle{#1}#1\ignorespaces}
\newcolumntype{C}{>{\bfseries}}

\let\oldtabular\tabular
\let\endoldtabular\endtabular
\renewenvironment{tabular}{\sffamily\oldtabular}{\endoldtabular}

\makeatletter
\renewcommand\bibsection{%
      \chapter*{\bibname\@mkboth{\bibname}{\bibname}}}
\makeatother

\AtBeginDocument{
  \DeclareSymbolFont{AMSb}{U}{msb}{m}{n}
  \DeclareSymbolFontAlphabet{\mathbb}{AMSb}}

\newcommand{\icode}[1]{{\lstset{basicstyle=\ttfamily\small}\lstinline@#1@}}

\definecolor{fixcolor}{rgb}{1,0.8,0.8}

\definecolor{commentcolor}{rgb}{0.8,1.0,0.8}

\title{Analysing Russian Trolls via NLP tools}
\author{Bokun Kong}
\date{\today}

\renewcommand{\thepage}{\roman{page}}

\makeindex
\begin{document}

%%%%%%%%%%%%%%%%%%%%%%%%%%%%%%%%%%%%%%%%%%%%%%%%%%%%%%%%%%%%%%%%%%%%%%%
%% Title page
\pagestyle{empty}
\thispagestyle{empty}
\input{titlepage}

%%%%%%%%%%%%%%%%%%%%%%%%%%%%%%%%%%%%%%%%%%%%%%%%%%%%%%%%%%%%%%%%%%%%%%%
%% Acknowledgements
\cleardoublepage
\pagestyle{empty}
\chapter*{Acknowledgments}
\addcontentsline{toc}{chapter}{Acknowledgments}
First and foremost, I would like to express my sincere gratitude to my advisor Dr. Dongwoo Kim, for his patience, motivation,  immense knowledge and his continuous support for my research. With his valuable help and comprehensive guidance in all stages of my project, I could accomplish this research and gain more knowledge about scientific reasearch on computer science.\\\\
Secondly, I would like to thank  Prof. Weifa Liang who are the course conveners for being supportive and providing significant help and advice throughout the entire period of this project.\\\\
Last but not least, I show gratitude to my family and friends who provide me with assistance and advices for this reseach project.

%%%%%%%%%%%%%%%%%%%%%%%%%%%%%%%%%%%%%%%%%%%%%%%%%%%%%%%%%%%%%%%%%%%%%%%
%% Abstract
\cleardoublepage
\pagestyle{headings}
\chapter*{Abstract}
\addcontentsline{toc}{chapter}{Abstract}
\vspace{-1em}
The fifty-eighth American presidential election in 2016 still arouse fierce controversy at present. A portion of politicians as well as medium  and voters believe that the Russian government interfered with the election of 2016 by controlling malicious social media accounts on twitter, such as trolls and bots accounts. Both of them will broadcast fake news, derail the conversations about election, and mislead people.\\\\ Therefore, this paper will focus on analysing some of the twitter dataset about the election of 2016 by using NLP methods and looking for some interesting patterns of whether the Russian government interfered with the election or not. We apply topic model on the given twitter dataset to extract some interesting topics and analyse the meaning, then we implement supervised topic model to retrieve the relationship between topics to category which is left troll or right troll, and analyse the pattern. Additionally, we will do sentiment analysis to analyse the attitude of the tweet. After extracting typical tweets from interesting topic, sentiment analysis offers the ability to know whether the tweet supports this topic or not. Based on comprehensive analysis and evaluation, we find interesting patterns of the dataset as well as some meaningful topics.

%%%%%%%%%%%%%%%%%%%%%%%%%%%%%%%%%%%%%%%%%%%%%%%%%%%%%%%%%%%%%%%%%%%%%%%
%% Table of contents
\cleardoublepage
\microtypesetup{protrusion=false}  % make sure numbers align in TOC
\pagestyle{headings}
\markboth{Contents}{Contents}
\tableofcontents
\listoffigures
\listoftables
\microtypesetup{protrusion=true}

%%%%%%%%%%%%%%%%%%%%%%%%%%%%%%%%%%%%%%%%%%%%%%%%%%%%%%%%%%%%%%%%%%%%%%
%% Here begins the main text
\mainmatter

%% Introduction
\chapter{Introduction}
\label{cha:intro}

Politically and socially relevant misconceptions, mis-information, and disinformation spread over social media such as Facebook and Twitter, posing a threat to democracy \citep{badawy2018analyzing}. The authors also mentioned that over the past decade, social media was witnessed to encourage democratic discourse in terms of social and political issues. As the population of users of social media increasing, the impact that electronic news media makes getting greater. Thus, despite significant potential to enable dissemination of factual information \citep{broniatowski2018weaponized}, people are more likely to be mis-leaded and mis-oriented by other users by reading and following their opinion \citep{mihaylov2015finding}. There are many opportunities for corporations, governments and other institutions to broadcast rumors, distribute misconceptions, and  manipulate public opinions by using other dishonest practices\citep{derczynski2014pheme}: from Occupy Wall Street movements \citep{conover2013geospatial, conover2013digital} to the Arab Spring \citep{gonzalez2011dynamics} and other civil protests \citep{gonzalez2013broadcasters, varol2014evolution}.\\\\Studies imply that 2016 U.S. Presidential election has been interfered by manipulating public opinion on social media such as twitter \citep{mihaylov2015finding, allcott2017social}. They also mentioned that the election has been intervened by online accounts that promoting divisive or conflicting issues on social or political aspect with aiming at manipulating public opinion. This type of online accounts is commonly known as trolls which basically is a person who deliberately provokes online conflict or distracts other users' opinions by broadcasting divisive and seditious content through social network. They aiming at  provoking others into an emotional response and derailing discussions \citep{buckels2014trolls}. Moreover, there are some other types of malicious social media accounts such as social bot which is an automated account run by an algorithm. It is considered to publish posts without human intervention, and it will also mislabel information and derail conversations. Although the troll and bot are different, they hold some similarities, and recent studies disclosed that the Russian government interfered the 2016 U.S. Election by using a combination of both bots and  trolls in social media \citep{badawy2018analyzing}.

\section{Problem Statement}
\label{sec:problemstatement}
As the intent of Russian trolls and bots are to deceive or create conflict during the election, this paper aims to analyse publicly available Russian troll dataset which active before and after the 2016 US Election on twitter. The dataset is published by Boatwright, Linvill, and Warren in 2018, involving approximately 3 million tweets from 2848 twitter accounts between February 2012 and May 2018. All the tweets collected in that dataset is highly likely to be controlled by the Russian government, and there are several types of Russian trolls such as left troll and right troll, and each type of troll has a certain behaviour and strategy.\\\\We focus on answering two basic research questions regarding the effects of the interference of Russian trolls on twitter in this paper.

\begin{enumerate}
    \item \textit{How to understand the behaviour of different types of Russian trolls?} We will analyse how different types of Russian trolls engaged with the 2016 US election, and how that may have helped propagate the Democratic Party of the United States and the Republican Party.
    \item \textit{How to understand the behaviour and strategy of Russian trolls change over time?} We will analyse how behaviours of different types of Russian trolls engaged with the 2016 US election changed from 2015 to 2017, and whether they manipulate public opinion effectively.
\end{enumerate}

\section{Motivations}
\label{sec:motivations}
In November 2017, Twitter released a list of 2,752 accounts, due to speculation about interference of Russian in the 2016 US presidential election via social media. Although Twitter did not specify how the dataset had been identified, they stated that these tweets are associated with the Internet Research Agency (RU-IRA) which basically is a troll farm from Russian that manipulate fake social media accounts derailing discussion about election and broadcasting divisive comments \citep{shane2017fake}. Twitter further mentions that some RU-IRA accounts "appear to have attempted to organize rallies and demonstrations, and several engaged in abusive behaviour and harassment" \citep{shane2017fake}. \\\\
We have little systematic evidence in terms of the impact of these accounts, how they affect the 2016 US election, and how they operate. We do not have a scientific classification framework to identify the role of trolls and a comprehensive understanding of how they affect the 2016 US Presidential Election before and after the election. Therefore, we compose this research report to introduce a sociological classification framework for the identification of roles for trolls, and the framework reveals their behaviours by conducting text based classification with topic retrieval as well as sentiment analysis. In the end, we will answer the two research questions above.

\section{Project Scope}
\label{sec:projectscope}
In this project, we perform textual topic model using Latent Dirichlet Allocation (LDA), which is a generative statistical model that explains sets of observations by unobserved groups explaining the reason why some sections of the data are similar \citep{blei2003latent}. Then, we will implement supervised topic model which is a statistical model of labelled documents \citep{mcauliffe2008supervised}, and the result of it will show the relation between retrieved topics and the predicted label representing whether supporting Liberal or Conservative. Finally, we do sentiment analysis to understand the sentiment carry out by specific tweets. All the text-based classification methods that we perform are in the context of natural language processing including plenty of subsets such as linguistics, text classification, natural language understanding, and information retrieval. Besides, this project involves some basic techniques of machine learning such as unsupervised learning and supervised learning. In this project, we stand on the shoulders of giants, making use of some tools for natural language processing such as The Natural Language Toolkit (NLTK) \citep{loper2002nltk} and Gensim \citep{rehurek_lrec} available publicly. NLTK is a platform  providing functions for applying statistical natural language processing (NLP) in natural language data, and many text processing libraries are contained for tokenization, classification, parsing, tagging, stemming, and semantic reasoning. Gensim is an open-source library for topic modelling and NLP with the help of statistical machine learning techniques, which designed to manage large text collections using incremental online algorithms and data streaming.

\section{Contributions}
\label{sec:contributions}
Findings from this research show that the majority of Russian trolls are promoting ideologically right by spreading pro-Trump material.Our findings presented in this work can be summarized as:\\
\begin{itemize}
	\item We offer a sociologically-grounded textual classification framework with the aim of identifying the role of trolls and their strategies by implementing text analysis to extract meaningful topics, and apply sentiment analysis to understand their attitude and standpoint.
	\item We analyse a new dataset about Russian trolls related to 2016 US presidential election, and make some interpretation of interesting pattern that we find in the new dataset.
	\item Our research indicates that Russian trolls are mostly promoting the Republican party, and spreading pro-Trump material.
\end{itemize}

\section{Report Outline}
\label{sec:outline}

There are 6 main chapters in this project:
\begin{itemize}
	\item \textbf{Chapter~\ref{cha:intro} “Introduction”} introduces the report, the motivation behind it, the problem that this project addressed, and its main contributions.
	\item\textbf{Chapter~\ref{cha:background} “Background and Related Work”} offers information about the background as well as related work about implementing text classification and sentiment analysis.
	\item\textbf{Chapter~\ref{cha:design} “Dataset” }describes the Russian troll dataset, and how we change the dataset in this work.
	\item\textbf{Chapter~\ref{cha:methodology}  “Methodology”} describes algorithms and technique for this project in detail, with explanation of the theory and how them work.
	\item\textbf{Chapter~\ref{cha:result} “Evaluation and Results” }compares different algorithms, illustrates the detailed model design and approaches, interprets interesting pattern.
	\item\textbf{Chapter~\ref{cha:discussion} “Discussion” }analyses the results, and answer questions that we proposed in Problem statement section.
	\item\textbf{Chapter~\ref{cha:conc} “Conclusion” }concludes this report and provide future work of this project.
\end{itemize}

%% Chapters
\chapter{Background and Related Work}
\label{cha:background}

As our study is to analyze Russian trolls on Twitter and to find out whether they interfered with 2016 U.S. Presidential Election, we are introducing related works in two aspects. Section~\ref{sec:social media} introduces public opinions of political trolls on social media in general, Section~\ref{sec:relatedwork} gives related works for analyzing political trolls on social media.\\

\section{Political trolls on social media}
\label{sec:social media}
Recently, social bots and online social trolls have attracted considerable academic attention. Studies have indicated that the impacts of trolls and social bots in social media are severe, they have influences on political discussions \citep{bessi2016social}, and the spread fake news \citep{shao2017spread}. In particularly, in political aspect, Donald Trump’s 2016 U.S. Presidential campaign has been supported by social media online trolls manipulated by Russian government \citep{flores2018mobilizing}. As described in Russian trolls dataset section, Russian trolls are generally have pro-Trump, conservative political agenda during 2016 U.S. election.\\\\
As social trolls have become progressively prevalent and influential, an increasing number of studies have been conducted to detect and analyse their roles in social media, and predict potential trolls \citep{cook2014twitter, davis2016botornot}. However, distinguishing and identifying specific types of trolls found to be difficult, because different categories of trolls could be controlled by the same person or group for a specific goal \citep{boatwright2018troll}. Their study indicates that each category of troll includes vastly various behaviours in terms of tweets content, reaction strategy to external events, and activity frequency patterns as well as volume, and their behavious will change over time.

\section{Related work}
\label{sec:relatedwork}

In this work, we aim to analyse behaviour of Russian troll dataset, and attempt to observe the strategy and roles of different types of trolls during and after the 2016 Election. By doing this, we use unsupervised learning method topic model to categorize tweets content according to content only, which can provide us with opportunities to observe and extract some meaningful topics to interpret their behaviours and strategies. We also implement supervised topic model to categorize tweets content along with prediction of possibilities of being a right troll or left troll. Additionally, we do sentiment analysis to conceive attitudes of representative tweets from a certain topic category, which observes whether trolls support the Democracy Party or the Republican Party. To our knowledge this has not been achieved for that Russian troll dataset in previous work. Some studies infer the political ideology of trolls using network-based methods \citep{badawy2018analyzing}, and temporal analysis of troll activity by analysing tweet volume as well as hashtag frequency in various time \citep{zannettou2019let}. Cosine distance and Levenshtein edit distance are used to calculate similarities between different trolls \citep{kumar2017army}, and time-sensitive semantic edit distance was proposed to improve performance of classification of roles of trolls based on their traces \citep{kim2019tracking}. Studies also compare left trolls and right trolls in terms of the number of re-tweets they received, number of followers, and the propaganda strategies they used by using machine learning method \citep{gorrell2019partisanship}. Moreover, machine learning model was proposed to predict troll accounts by considering bot likelihood, political ideologies and activity-related account which interacted with the trolls by sharing their contents \citep{badawy2018analyzing}.\\\\

\section{Summary}

In this chapter, we introduced the public opinions about political  trolls on social media, in particular Russian trolls on Twitter, and then we summarized the state-of-art techniques done by other studies in order to analysing Russian trolls, and then, mentioned how we analyse Russian troll dataset in our research. After introducing the Background, we will introduce the Russian troll dataset collected from Twitter, how we use the dataset in this work, and then we move on to introduce our design and implementation in detail.

\chapter{Dataset}
\label{cha:design}

In this chapter, we introduce the Russian troll dataset that we found and modifications that we made, and we also explain content of the dataset in detail.

\section{the Russian trolls Twitter data set}
\label{sec:des-hotpath}

In this project, we use a publicly accessible  dataset\footnote{Link to the dataset: https://github.com/fivethirtyeight/russian-troll-tweets/} containing verified Russian troll activities on Twitter, published by Clemson University researchers \citep{badawy2018analyzing}. The complete Russian troll dataset contains 2848 Twitter handles, nearly 3 million tweets in total. The dataset is considered to be the most up-to-date and comprehensive record of Russian troll activities on social media. The time range of included tweets were posted between February 2012 and May 2018, most between 2015 and 2017.\\\\
In order to understand behaviours of the Russian trolls, we aim to distinguish different types of trolls and relative topics based on their authored text and publish dates. Since each handle might has multiple types of Russian trolls categories broadcasting different types of news, we aim to analyse contents of tweets and attempt to detect the Russian trolls category for each tweet, rather than detecting the Russian trolls category for each handle. The Clemson researchers categorize multiple types of Russian trolls: left troll; right troll; news feed; hashtag gamer; and fearmonger. In this work, we focus on two main parts: analysing Russian troll dataset using topic model; analysing Russian troll dataset using supervised topic model combined with sentiment analysis. For analyzing the dataset using both topic model and supervised topic model, we involve the top 2 most frequent trolls: left troll,and right troll. In addition, we only consider tweets content in English, and we do not analyse tweets in other language or using other symbols.\\\\
Various types of Russian troll category have different bahaviours and roles, mentioned by Boatwright, Linvill, and Warren in 2018. News feeds trolls inclined to magnify, contribute to public panic and disorder, and pretend to be legitimate local news aggregators. While news feeds trolls have no specific political inclination, the purposes of left trolls and right trolls are clear. The right trolls represent Republican Party having the slogan that “Make American Great Again”. Acronym “MAGA” was used by Donald Trump during his election campaign in 2016, and it is also the central theme of his presidency. The right trolls behave like typical Trump supporters by broadcasting news that benefit his election. On the contrary, left trolls did not support the Democratic Party, they tend to divide liberal party against itself and result in  lower voter turnout. They act derisively towards Hillary Clinton, and attempt to mimic Black Lives Matter activities, supporting Bernie Sanders who was the alternative Democrat Presidential Nominee. In general, Russian trolls have a strong political inclination to support Donald Trump, rather than Hillary Clinton. We will analyse the dataset and prove this strategy is actually the goal of Russian trolls in this research.\\\\

\section{Modification of the dataset}
\label{sec:mod}

The Russian troll dataset involves a huge number of tweets as well as corresponding account information in a variety of different languages. In this work, we focus on analysing tweets in English, which are majority in the dataset, thus, we pre-process the dataset to drop tweets with “language” is not “English”. And we only use tweets with “account\_category” is “RightTroll” or “LeftTroll” in order to find strategy of left and right trolls. The number of tweets in the dataset decrease from 2435342 to 984045. As we can see, the number of right trolls is almost twice than the number of left trolls.\\

\begin{table*}[h]
	\centering
	
	\caption{Description of the original dataset.}
	
	\label{tab:t1}
	% Every table you want to use the bold macros in must start with the B specifier.
% The C specifier makes the corresponding column bold. (so starting with "BCr"
% makes the first column bold and right-aligned)
% You must insert ^ between each column specifier.
% Then, the \rowstyle{\bfseries} turns that row bold.

\begin{tabular}{BCr^r^r^r}
\toprule
\rowstyle{\bfseries}
       & Left trolls  & Right trolls   & Total number\\
\midrule 
Number              & 984045        & 369313         & 614732     \\

\bottomrule
\end{tabular}

%%% Local Variables: 
%%% mode: latex
%%% TeX-master: t
%%% End: 

\end{table*}

Moreover, we restrict the time range of tweets publish dates in one year: 2015, 2016, and 2017. \\

\begin{table*}[h]
	\centering
	
	\caption{Description of dataset in different time range.}
	
	\label{tab:t2}
	% Every table you want to use the bold macros in must start with the B specifier.
% The C specifier makes the corresponding column bold. (so starting with "BCr"
% makes the first column bold and right-aligned)
% You must insert ^ between each column specifier.
% Then, the \rowstyle{\bfseries} turns that row bold.

\begin{tabular}{BCr^r^r^r}
\toprule
\rowstyle{\bfseries}
 Year      & Left trolls  & Right trolls   & Total number\\
\midrule 
2015	&31439		&127337		&158776		\\
2016	&184949		&145352		&330301		\\
2017	&147771		&339293		&487064		\\
\bottomrule
\end{tabular}

%%% Local Variables: 
%%% mode: latex
%%% TeX-master: t
%%% End: 

\end{table*}

We notice that the number of trolls is increasing from 2015 to 2017. By restricting the time range, we are able to observe strategies of different types of the Russian trolls and how strategies change over time more precisely.\\

\section{Summary}
After we get a thorough understanding of the dataset, let’s focus on methodology that how to analyse the dataset to observe their intends and strategies.\\

\chapter{Methodology}
\label{cha:methodology}

In our research, we analyse the Russian troll dataset using several natural language processing techniques, which can be separated to three main parts: Section~\ref{sec:lda}topic model, Section~\ref{sec:slda}supervised topic model and Section~\ref{sec:sa}sentiment analysis. Before that, we do Section~\ref{sec:pre}data pre-processing procedure at first.

\section{Pre-processing}
\label{sec:pre}
In order to analysing the Russian troll dataset using topic model and supervised topic model, data pre-process is needed. After transforming raw data into an understandable format, we can easily train the dataset.\\\\
In this section, we introduce a great variety of data preprocessing methods that we used for both topic model and supervised topic model. All of them play an important role in the process of lexical analysis.

\subsection{Tokenize}

The first procedure that we take is tokenizing the tweets dataset, we record each word as a token and store into a dictionary. More specifically, tokenization breaks up a sequence of strings into pieces such as words, symbols and other elements. We use a function called \icode{WordPunctTokenizer} in NLTK library for splitting strings into tokens. The function is based on the regular expression \icode{\\w+|[\^\\w\\s]+}, which tokenize a text into non-alphabetic and alphabetic characters with a sequence. For example, a sentence \textit{"Good muffins cost \$3.88 in New York. Please buy me two of them. Thanks."} has tokens: ['Good', 'muffins', 'cost', '\$', '3', '.', '88', 'in', 'New', 'York', '.', 'Please', 'buy', 'me', 'two', 'of', 'them', '.', 'Thanks', '.'] by using that function. Some of the tokens are punctuation or non-alphabetical characters which are meaningless, thus, we remove punctuation and non-alphabetic characters.\\

\subsection{Modify tokens}
Now all tokens are alphabetic characters, but those tokens still cannot be used directly, since some tokens are still meaningless or some frequent tokens even pose negative effects to the topic model. In order to solve these problems, we first convert all tokens in lower cases. Next, we remove stop-words which refers to the most common words in a specific language such as “a”, “the”, “is” which are meaningless to the model to analyse the tweets content, and stop words should be removed before processing of natural language data. In this work, we make use of function from Genism library based on English. Besides, we filter out tokens that have length equal to 1 or greater than 14, since these tokens might be useless or not even a word in English.\\\\
After removing useless tokens, we should now consider that a word may have various forms. For example, a verb has past participle and past tense, a none has its plural and so on. Thus, we turn tokens into lemmas and stems to summarise general words with similar meanings. For example, we change token \texttt{“better”} to \texttt{“good”} for lemmatization, and \texttt{“running”} to \texttt{“run”} for stemming. Now tokens have been pre-processed and can be trained in topic model. \\

\section{Topic model}
\label{sec:lda}
In fields of machine learning and natural language processing, topic model is a type of statistical model for observing the abstract "topics" that appear in a collection of documents. We use topic model in our work because we would like to extract “topics” from a huge number of tweets in the Russian troll dataset. By using that data mining tool for observation of hidden semantic structures in the Russian troll text body dataset. The "topics" produced by topic modelling techniques are clusters of similar words.\\\\

\subsection{Theoretical model}
An early topic model called probabilistic latent semantic analysis (PLSA), was created by Thomas Hofmann in 1999 \citep{hofmann1999probabilistic}. The most commonly used topic model is Latent Dirichlet allocation (LDA) which is generative probabilistic model, and a generalization of PLSA, developed by David Blei, Andrew Ng, and Michael I. Jordan in 2003. LDA introduces sparse Dirichlet prior distributions over document-topic and topic-word distributions, which contains the intuition that a small number of topics is covered by documents and that topics often use a small number of words \citep{blei2003latent}.\\\\
In this work, we apply Latent Dirichlet allocation (LDA) to the Russian troll tweets content and split them into topics. Each tweet contains a number of words, and tweet is regarded as document in LDA. In LDA, each document can be thought as a mixture of various topics that are assigned to it via LDA. It is identical to PLSA, except that the topic distribution is assumed to have a sparse Dirichlet prior in LDA. The intuition of sparse Dirichlet is that the document covers only a small set of the topics and those topics only use few words frequently, which contribute to better words ambiguity elimination and more precise assignment of documents to topics \citep{girolami2003equivalence}. Given the context of topic modeling, it can be assumed that all the documents are generated by some hidden, or latent variables (topics, topic assignments and topic proportions). In order to generate these latent factors, we use statistical inference. Specifically, approximate posterior inference methods such as Gibbs sampling and variational inference.\\\\
In LDA, the topic distribution for each document is distributed as:

$$\displaystyle \theta \sim \operatorname {\textbf{Dirichlet}} (\alpha)$$\\
Where \textbf{Dirichlet}($\alpha$) denotes the Dirichlet distribution for parameter $\alpha$. The term distribution  is also modeled by a Dirichlet distribution, but under a different parameter $\beta$.

$$\displaystyle \varphi \sim \operatorname {\textbf{Dirichlet}} (\beta)$$\\
The final goal of LDA is to estimate the  $\theta$ and $\varphi$ . The basic idea is the higher the value $\alpha$ denotes that each document is likely to contain a mixture of most of the topics, and the higher the value $\beta$ the more likely each topic containing a mixture of the majority of words.\\\\
\begin{figure}[t!]
	\centering
	\includegraphics[width=0.6\textwidth]{{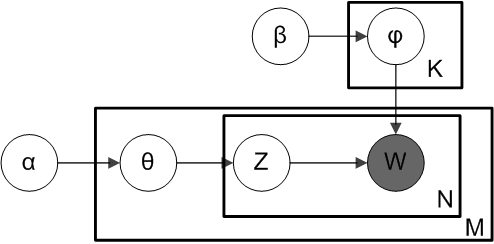}}
	\caption{Plate notation for LDA with Dirichlet-distributed topic-word distributions.}
	\label{fig:LDA}
\end{figure}
The LDA model can be demonstrated in plate notation in figure~\ref{fig:LDA} , which concisely captures the dependencies among these variables, and the boxes are "plates" representing repeated entities. The outer plate represents documents, while the inner plate denotes the repeated word positions. M refers to the number of documents, N is the number of words in a document, and the variable names are defined as follows:

\begin{itemize}
	\item $\alpha$ is the parameter of the Dirichlet prior on the per-document topic distributions
	\item $\beta$ is the parameter of the Dirichlet prior on the per-topic word distribution
	\item $\theta_{i}$ is the topic distribution for document i
	\item $\varphi _{k}$ is the word distribution for topic k
	\item $z_{ij}$is the topic for the j-th word in document i
	\item $w_{ij}$ is the specific word
	
\end{itemize}
The grayed W means that words $w_{ij}$ are the only observable variables, and the other variables are latent variables. Based on generative process [David, Andrew, and Michael, 2003], each document (w) in a corpus (D) is generated by:\\

\begin{enumerate}
\item Choose  $\displaystyle \theta _{i}\sim \operatorname {Dirichlet} (\alpha )$, where  $i\in \{1,\dots ,M\}$ and Dirichlet($\alpha$) is a Dirichlet distribution with a symmetric parameter $\alpha$ which typically is sparse ($\alpha$ < 1)
\item Choose  $\displaystyle \varphi _{k}\sim \operatorname {Dirichlet} (\beta )$, where  $k\in \{1,\dots ,K\}$ and $\beta$  typically is sparse

\item For each of the word positions \textit{i}, \textit{j}, where $i\in \{1,\dots ,M\}$ , and $j\in \{1,\dots ,N_{i}\}$ 
\begin{enumerate}
    \item Choose a topic ${\displaystyle z_{i,j}\sim \operatorname {Multinomial} (\theta _{i})}$
    \item Choose a word $w_{i,j}$ from $p(w_{i,j}|z_{i,j}, \beta)$, a multinomial probability conditioned on the topic $z_{i,j}$
\end{enumerate}
\end{enumerate}
$\theta$ represents a random distribution Dirichlet parameterized by a vector of length K $\alpha$. K being the number of topics that have been selected. N is a fixed vocabulary of words $w_{i,j}$ Then for 1..N, selecting a topic $z_{i,j}$ taken from a discrete (multinomial) distribution parameterized by $\theta$, and then select a word based on a probability conditioned on the topic selected before.\\\\
Given a corpus, only variable $w_{i,j}$ can be observed, $\alpha$,  $\beta$ are prior parameters, and $w_{i,j}$, $\theta_{i}$ and $\varphi_{i}$ are latent variables that need to be estimated based on the observed variables. The inference is based on probability graph model including exact inference and approximate inference. Exact inference is hard to implement in LDA; thus, people commonly use approximate inference to learn latent variables in LDA such as Gibbs Sampling algorithm. In short, it works by sampling each of those variables given the other variables.\\\\
Gibbs Sampling is a special case of Markov-Chain Monte Carlo algorithm. The algorithm works by selecting one dimension of the probability vector each time, giving the value of the variable of other dimensions, sampling the value of the current dimension, and iterating until the output of the parameter to be estimated is converged.\\\\
\begin{figure}[h]
	\centering
	
	\includegraphics[width=\textwidth]{{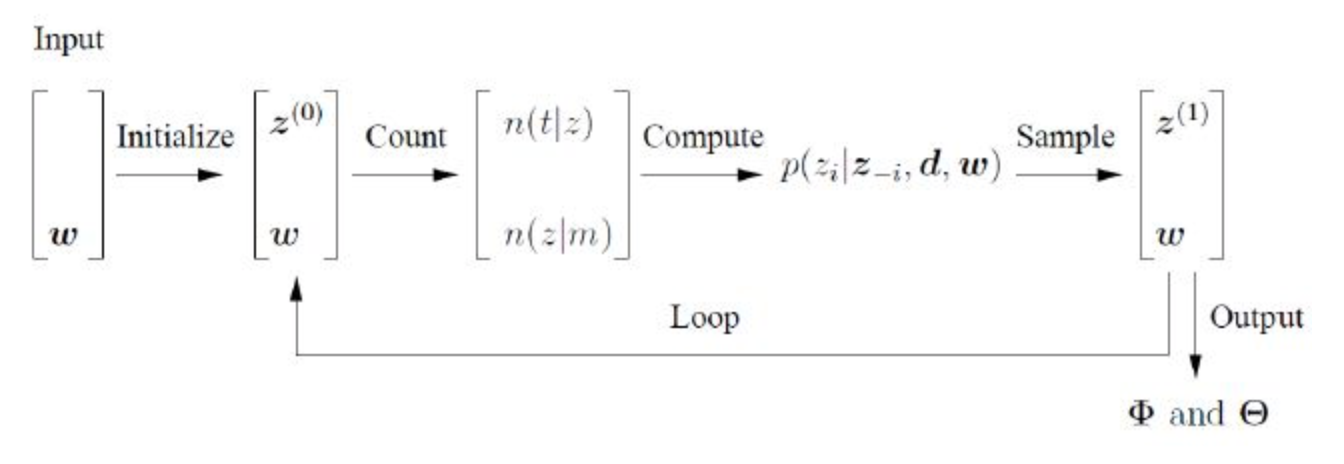}}
	\caption{The procedure of learning LDA by Gibbs sampling.}
	\label{fig:gibbs}
\end{figure}

Figure~\ref{fig:gibbs} represents procedure of learning LDA using Gibbs sampling \citep{wang2008distributed}. Initially, we should assign a topic $z^{0}$ to each word in the text at random, and then count the number of term $t$ appearing under each topic $z$ and the number of words in topic $z$ appearing under each document $m$, compute $p(z_{i}|z_{-i},d ,w)$ for each iteration. After estimating the probability of assigning each topic to the current word based on the topic assignment of all other words, we sample a new topic $z^{1}$ for the word according to this probability distribution. 

\subsection{Implementation}
In this research, we implement topic model on the Russian troll dataset by using LDA. The Gensim library provides useful models for LDA and other topic model algorithms, as well as some data pre-processing functions, and comprehensive corpora for natural language processing. In order to train LDA, we first create a lexical dictionary of corpus, assigning an index to each individual word, then we create term-document matrix by converting document into the bag-of-words (BoW) format that contains a list of token id and token count. In the end, we train the LDA model imported from Gensim by providing term-document matrix and the dictionary we created before, along with other parameters such as the number of iterations, decay, and offset. Besides, we also need to determine the number of topics. It’s significant to note that the algorithm makes no determination of what the topic should be or that the number of topics is adequate for the corpus. Therefore, human intuition and mathematical optimization should be applied to determine suitable number of topics. \\\\
We improved our model by filter out the most common words and least common words. Some frequent words in the corpus may still be useless, as those words might be general, and have no specific meaning. Additionally, some rare words might still be meaningless, since those words are too specific, and are not mentioned in other documents; thus, those words will not generate a good topic. The number of words that should be filtered out depends on the size of the dictionary, and a large dictionary should filter out more words. We deliver experiments on comparing parameter settings in Evaluation and Result chapter. \\\\
Moreover, we modify the term-document matrix by adding term frequency–inverse document frequency (tf-idf) technique, which will lead to better performance on generating topics from LDA. Tf-idf is a numerical statistic intending to reflect the importance of a word to a document in corpus. The tf–idf value increases in proportion to the number of times a word appears in the document and is offset by the number of documents in the corpus containing the word, which mitigates the negative effect that certain words usually appear more frequently. Tf-idf is composed of two components: term frequency which means the number of times a term occurs in a document; and inverse document frequency. Let document frequency $df_{t}$ be the number of documents in the collection that contain a term $t$. The idf can be defined as:

$$idf_{t} = log(\dfrac{N}{df_{t}})$$\\
where N is the total number of documents.

\section{Supervised topic model}
\label{sec:slda}
The LDA that we applied above is an unsupervised technique, but in this section, we attempt to convert hidden semantic structure to be used in a supervised classification problem, which means implementation of supervised latent Dirichlet allocation (sLDA), a statistical model of labelled documents. The model accepts various response types with the goal of inferring latent topics . Given an unlabelled document, we first infer its topic structure using a fitted model, then form its prediction.\\\\
Previously, unsupervised LDA was used to construct features for classification. Hopefully that LDA topics are useful for categorization, as they reduce the dimension of data. However, experiments indicate that fitting unsupervised topics might not be an ideal option when the goal is prediction \citep{blei2003latent}. In text analysis, joint topic model for words and categories was developed by \citep{mccallum2006multi}, and LDA model for the prediction of caption words from images is developed by \citep{blei2003modeling}. Moreover, “labelled LDA” was proposed for genes and protein function categories, and “labelled LDA” is also a joint topic model \citep{flaherty2005latent}.\\\\

\subsection{Theoretical model}
In supervised latent Dirichlet allocation (sLDA), a response variable associated with each document is added to LDA, and the variable might be numerical labels such as stars given to a movie, or either right troll or left troll represented by a number \citep{mcauliffe2008supervised}. The sLDA models the responses and the documents together, thus, it will find latent topics that best predict the response variables for future unlabelled documents. \\\\
For the model parameters: $K$ topics $\beta_{1:K}$ where each $\beta_{k}$ a vector of term probabilities), a Dirichlet parameter $\alpha$, and response parameters $\eta$ and $\delta$. In the sLDA model, document and response results from the following generative process:\\
\begin{enumerate}
\item Draw topic proportions $\theta | \alpha$ $\sim$ $\operatorname {Dirichlet} (\alpha).$
\item For each word

(a) Draw topic assignment $z_{n} | \theta$ $\sim$ $\operatorname {Multinomial} (\theta).$

(b) Draw word $w_{n} | z_{n}$, $\beta_{1:K}$ $\sim$ $\operatorname {Multinomial} (\beta_{z_{n}}).$
\item Draw response variable $y|z_{1:N}, \eta, \delta$ $\sim$ $\operatorname {GLM} (\bar{z}, \eta, \delta).$, where we define
$$\bar{z} := (1/N ) \sum_{n=1}^{N} z_{n}.$$
\end{enumerate}
\begin{figure}[h]
	\centering
	
	\includegraphics[width=0.6\textwidth]{{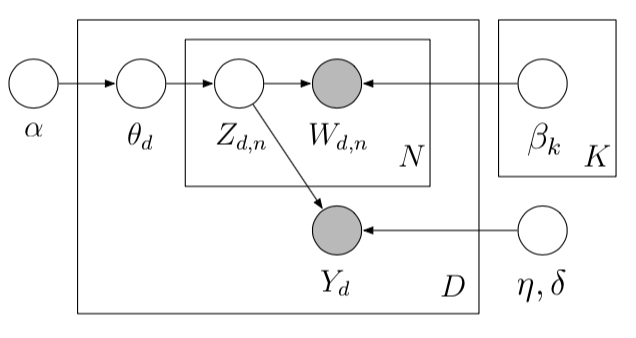}}
	\caption{The graphical model representation of supervised latent Dirichlet allocation (sLDA).}
	\label{fig:slda}
\end{figure}
Figure~\ref{fig:slda} illustrates the probability distributions for  this generative process in a plate notation. All the nodes are random variables, and edges denote possible dependence. The observed variables are shaded nodes, while hidden variables are unshaded nodes.\\\\
The  distribution of the response is a generalized linear model developed by \citep{mccullagh1989generalized}. There are two main components in GLM: "random component" and "systematic component":
$$p(y|z_{1:N}, \eta, \delta) = h(y, \delta) \exp{\dfrac{(\eta^{T}\bar{z}) - A(\eta^{T}\bar{z})}{\delta}}$$
$\eta^{T}\bar{z}$ is a natural parameter and $\delta$ is dispersion parameter. This equation is an exponential family, with base measure $h(y, \delta)$, sufficient statistic $y$, and log-normalizer $A(\eta_{T}\bar{z})$.\\\\
The sLDA is different from general GLM that the covariates for sLDA are the unobserved empirical frequencies of the topics in the document. The words and the response are combined, as latent variables determine the words of the document, in the generative process. In general, sLDA is similar with LDA with an additional prediction for the label. In order to analyse data with sLDA, three computational problems need to be addressed. Firstly,  posterior inference, the conditional distribution of the latent variables given model parameters and its words should be computed. The computation of distribution is the same as computing distributions in LDA, which is difficult to compute. There are two ways to approximate it: Gibbs sampling and variational inference, as in LDA in topic model, we choose to use Gibbs sampling to infer distribution. Second is parameter estimation, and it similar with LDA that estimates the Dirichlet parameters $\alpha$, and topic multinomials $\beta_{1:K}$, but in sLDA, GLM parameters $\eta$ and $\delta$ are also be estimated. Maximum likelihood estimation is implemented based on variational expectation-maximization. Finally, prediction. sLDA predicts a response $y$ from a newly observed document and model parameters based on $E[y|w_{1:N}, \alpha, \beta_{1:K}, \eta, \delta]$. The expectation can be approximated by $E_{q}[\mu(\eta^{T}\bar{Z})]$.

\subsection{Implementation}
In this research, we implement supervised topic model on the Russian troll dataset by using sLDA. We implement sLDA based on supervised topic model tutorial analysing movie dataset available online\footnote{Link to github: https://github.com/dongwookim-ml/python-topic-model}. The tutorial of implementing sLDA was posted on Github publicly, and we use the basic idea of algorithm to train sLDA on our Russian troll dataset. In our Russian troll dataset, the prediction is the label either right troll or left troll. In the dataset, the label is string “leftTroll” or “rightTroll”, thus, we convert them into numeric labels that left troll equals to -1 and right troll equals to 1. This kind of document-response corpora that tweets content with numerical labels could be trained in sLDA model easily. As sLDA is a supervised learning technique, it can predict unlabelled data after training by labelled training dataset, and we can evaluate the performance of the sLDA algorithm on test dataset. Therefore, we separate the Russian troll dataset to training set consisting 70\% of dataset, and 30\% of testing dataset for evaluation. The testing dataset also called held-out documents. \\\\
We train supervised LDA with Gaussian response variables where the coefficient parameter of Gaussian distribution is the predicted label. In sLDA, we first initialize topics to each document randomly. Then we use Stochastic Expectation Maximisation algorithm to train the sLDA model with the number of max iteration 100. Additionally, we compute mean absolute error and log likelihood to evaluate the performance of algorithm. For evaluation on held-out documents, we fit the test corpus into trained sLDA model, and the algorithm learns parameters from held-out documents and then generates held-out documents topic assignments. After normalising topic assignments, the predicted label is obtained by dot producing normalised topic assignments and coefficient parameter of Gaussian distribution of topic labels.\\\\
Apart from that, we separate the dataset to further analyse the Russian troll with restricted time range. We separate dataset into three parts depending on publishing date: tweets published in 2015, 2016 and 2017.By analysing the Russian troll with restricted time range, we might observe strategies of Russian troll in different time range and how they change over time. For the implementation, we simply separate dataset into three parts for both training set and testing set, and then train the sLDA model correspondingly with time restricted datasets.\\\\

\section{Sentiment analysis}
\label{sec:sa}

Sentiment analysis denotes studying, identifying, extracting subjective information, emotions from texts, and classifying opinions as negative, positive or neutral using a variety of natural language processing techniques such as text analysis, and computational linguistics. Sentiment analysis can be applied at different levels of scopes, for example, document level focuses on the sentiment in a complete document or paragraph; Sentence level analyses sentiment in a single sentence; Sub-sentence level obtains the sentiment in sub-expressions in a sentence. There are various types of sentiment analysis and sentiment analysis tools. For example, Fine-grained Sentiment Analysis focuses on polarity considering categories: very positive, positive, neutral, negative, and very negative. Emotion detection observe feelings and emotions (e.g. angry, happy, sad, etc.). Intent analysis identifies intentions for example interested or not interested with given text.\\\\
\subsection{Theoretical model}
In terms of sentiment analysis algorithms, there are various approaches and algorithms to implement sentiment analysis systems, which can be concluded in three categories:
\begin{enumerate}
	\item \textbf{Machine Learning:} \\This approach relies on machine-learning techniques and diverse features to construct a classifier that can identify sentiment expressed from text.
	\item \textbf{Lexicon-based:} \\This method uses a great number of words annotated by polarity score, to decide the general assessment score of a given content.
	\item \textbf{Hybrid:} \\Hybrid algorithm combines both machine learning and lexicon-based approaches.
\end{enumerate}
\begin{figure}[h]
	\centering
	
	\includegraphics[width=\textwidth]{{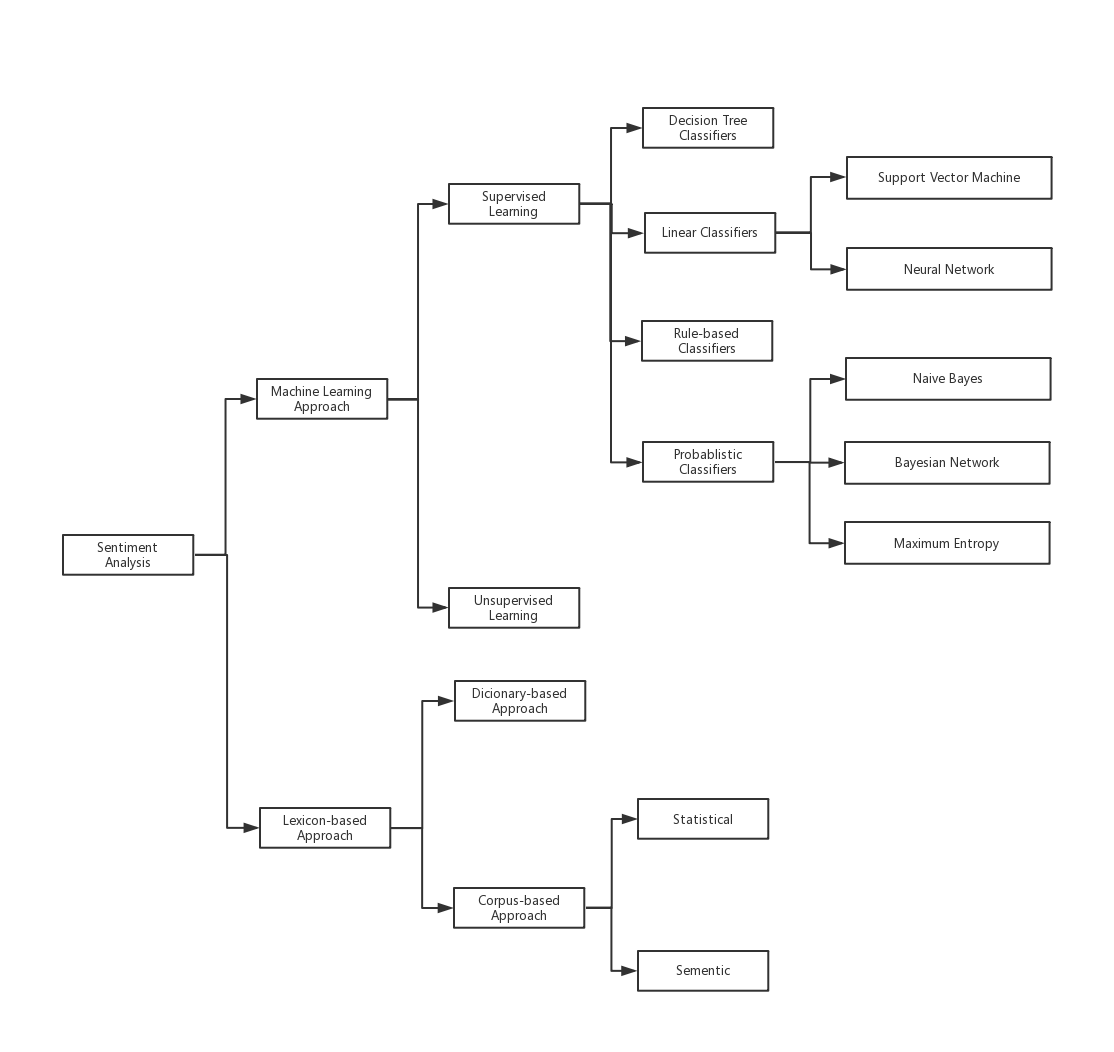}}
	\caption{Sentiment classification techniques.}
	\label{fig:sa}
\end{figure}
Figure~\ref{fig:sa} displays some sentiment classification techniques. Lexicon-based algorithms are very naïve since they do not consider the sequences of combinations of words which would convey different sentiment. Therefore, we focus on machine learning sentiment classifications in this research.\\
\begin{figure}[h]
	\centering
	
	\includegraphics[width=\textwidth]{{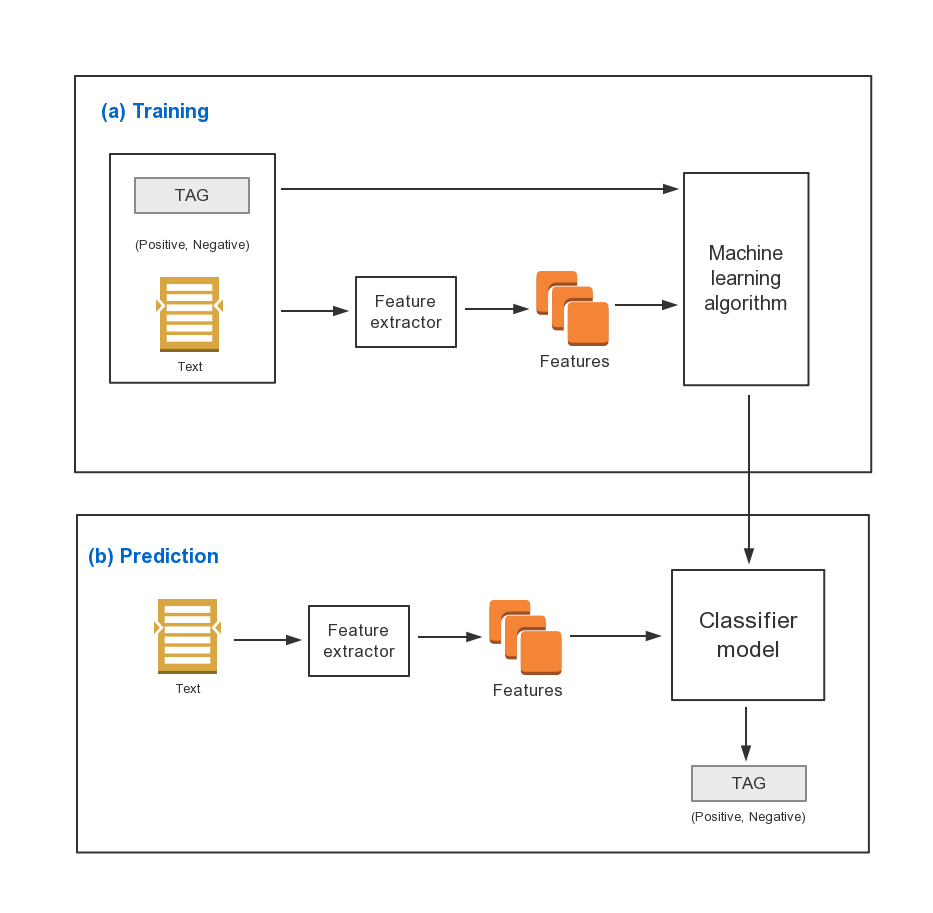}}
	\caption{Sentiment analysis procedures.}
	\label{fig:saml}
\end{figure}\\
Figure~\ref{fig:saml} illustrates procedures of Sentiment analysis. An important step is to transform the text into a numerical representation such as vector, containing frequencies of a words. There are many approaches to achieve feature extraction: bag-of-words or bag-of-ngrams with their frequency are commonly used, and new extraction techniques based on word embedding are also popular. For classification algorithms, there are many popular statistic models such as Naïve Bayes, Logistic Regression, Support Vector Machines, and Neural Networks.

\subsection{Implementation}
By only using supervised topic model or topic model, we cannot identify troll’s opinion from this unstructured information. Sentiment analysis clarifies the polarity of either left troll or right troll that whether they conceive positive or negative opinion, which conveys significant information for analyzing of the strategies of the Russian trolls. In this study, we regard sentiment classification as a binary problem involving positive and negative opinions. \\\\
As the original Russian troll dataset does not contain sentiment category for tweets, thus, we cannot train the sentiment classification model based on the Russian troll dataset. We, therefore, make use of other datasets containing sentiment labels that publicly available on the Internet. Our intuition is that by training the sentiment classifier on other dataset, we can use the classifier to predict sentiment of tweets in the Russian troll dataset. Initially, we trained the model with a movie review dataset with sentiment labels. We then notice that we should train a dataset which is similar to the Russian troll dataset as much as possible to obtain better performance of our sentiment classifier. We choose to use a Twitter dataset containing 1.6 million tweets provided on kaggle\footnote{Link to kaggle dataset: https://www.kaggle.com/kazanova/sentiment140}, and this dataset is better than the movie review dataset, since data is extracted from Twitter, which is similar to our testing dataset. We create lists to store positive words and negative words separately, and we only use approximately 10\% of dataset, as the size of it is large. The 5000 most frequent words are used to extract features. After creating features for each tweet, we separate feature sets to two parts: training set and testing set, then we train the classifier with training set.
\\\\
We construct various of classifiers using different algorithms such as Naïve Bayes and Logistic Regression, and compare them by classifier accuracy and F1 score. Then we ensemble these classifiers into one model, developing a simple measurement the degree of confidence in the classification based on majority of votes of these classifiers. The final ensemble sentiment classification model is used to predict polarity of the Russian troll dataset. We only analyse interesting topics with extremely high or low predicted value, which means those topics are more likely to have left troll or right troll inclination. Representative tweets for interesting topics are retrieved for sentiment analysis by selecting relatively high normalized document-topic matrix value for each document.

\section{Summary}
In this chapter, we introduce the design of our methods for analysing the Russian troll dataset. We explain theories of the algorithms that we use in detail, and demonstrate how to implement these algorithms with the Russian troll dataset. Firstly, Latent Dirichlet allocation is applied for analysing topic model of the dataset. And then we implement supervised Latent Dirichlet allocation (sLDA) to predict types of troll categories for each topic. Additionally, we restrict the time range of tweets in the dataset, and then apply sLDA to further analyse how trolls change their behaviours over time. Finally, we do sentiment analysis to predict the polarity of tweets from interested topics. The attitudes of troll in each predicted topic offer a holistic view of strategies of Russian trolls. In the next chapter, we will present the results of our proposed method and evaluate results.

%%% Local Variables: 
%%% mode: latex
%%% TeX-master: "paper"
%%% End: 

\chapter{Evaluation and Results}
\label{cha:result}
We report our staged results in this chapter and evaluate results with different parameter settings in our implemented algorithms. Specifically, the evaluation includes three parts: Chapter~\ref{sec:etm} evaluates performance of the Latent Dirichlet allocation (LDA) for topic model on the Russian troll dataset by recognizing whether the algorithm generates meaningful topics qualitatively. Chapter~\ref{sec:estm} evaluates results of supervised Latent Dirichlet allocation (sLDA) performed on the dataset by interpreting meaningful topics with extreme predicted label values representing high possibilities to be right or left troll manually. In chapter~\ref{sec:esa}, we compare the performance of different sentiment classifiers by accuracy and F1 measurement.\\\\
In evaluating topic model and supervised topic model, we noticed that distinguishing meaningful topic is more based on human intuition, and we do not generate a mathematical way to categorize the usefulness of a constructed topic automatically. We can compare the performance of algorithm by calculating percentage of number of meaningful themes generated over the total number of topics, and the percentage can be regarded as accuracy of the model. Therefore, we develop a quantitative method to evaluate the LDA that calculates the percentage of meaningful topics

\section{Topic model}
\label{sec:etm}
As mentioned in Methodology chapter, we first implement LDA to analyse topic model using Bag of Words, we do not filter out frequent words, and we only keep 100,000 most frequent words, since other words are too rare and useless. Note that the algorithm does not determine the appropriate number of topics for the input. Therefore, human intuition should be applied to determine suitable number of topics. Thus, we assign the number of topics to a moderate number initially which is 20. \\
\begin{table*}[h]
	\centering
	\caption{A useless topic (These words do not share any coherent theme).}
	\label{tab:1n}
	
\resizebox{\textwidth}{10mm}{
    \begin{tabular}{BCr^r^r^r^r^r^r^r^r^r^r}
    \toprule
    \rowstyle{\bfseries}
      & w1    & w2    & w3    & w4    & w5    & w6      & w7    & w8    & w9          & w10  \\
    \midrule 
    Words                  & step  & body  & want  & amp   & http  & protect & use   & know  & information & like         \\
    Probabilities          & 0.016 & 0.013 & 0.012 & 0.010 & 0.009 & 0.008   & 0.008 & 0.007 & 0.007       & 0.007 \\
    
    \bottomrule
    \end{tabular}
}
\end{table*}\\
Table~\ref{tab:1n} shows a useless topic generated from the model. Note that we only list 10 words with high probabilities for each topic, and  the number in front of each word is probability that the word will be selected after the topic has been selected in the process of generating texts. In this topic, we find some words such as "want", "like", and "use" are frequent words and they have no meaning in the theme. Word like "http" is proper noun, and frequent proper nouns do not have meaning neither. Therefore, we cannot generate a theme from those words in that topic. However, that model generate good topic such as table~\ref{tab:1p}. \\
\begin{table*}[h]
	\centering
	\caption{A useful topic (Containing theme "Islam and War").}
	\label{tab:1p}
	\resizebox{\textwidth}{10mm}{
    \begin{tabular}{BCr^r^r^r^r^r^r^r^r^r^r}
    \toprule
    \rowstyle{\bfseries}
      & w1    & w2    & w3    & w4    & w5    & w6      & w7    & w8    & w9          & w10  \\
    \midrule 
    Words                  & refugees  & christmas  & islamkills  & brussels   & rich  & moment & syrian   & tip  & trump & doctor         \\
    Probabilities          & 0.022 & 0.020 & 0.017 & 0.013 & 0.012 & 0.010   & 0.009 & 0.008 & 0.008       & 0.007 \\
    
    \bottomrule
    \end{tabular}
}
\end{table*}\\
This figure is a meaningful topic, since it contains words such as "refugees", "islamkills", and "syrian", and the probability of word "refugees" is high. That topic can be concluded in the theme of \textit{Islam and War}. We record the number of meaningful topic with default parameter setting is 2 and the total number of topic is 20, thus, it equals to percentage 10\%. Next, we remove some frequent words from the dictionary that we created, and we also remove some rare word by guaranteeing all words appear in more than 5 documents. Moreover, we apply term frequency–inverse document frequency (tf-idf) technique in the LDA model to adjust for the fact that some words appear more frequently in general. The modified algorithm generates many interesting topics which displayed in table~\ref{tab:2p}. \\
\begin{table*}[h]
    \begin{subtable}[b]{\textwidth}
    	\centering
    	\caption{Supporting Hillary.}
    	\label{tab:2p1}
    	\resizebox{\textwidth}{10mm}{
    \begin{tabular}{BCr^r^r^r^r^r^r^r^r^r^r}
    \toprule
    \rowstyle{\bfseries}
      & w1    & w2    & w3    & w4    & w5    & w6      & w7    & w8    & w9          & w10  \\
    \midrule 
    Words                  & hillaryclinton  & remember  & great  & poll   & job  & cnn & truth   & mayor  & god & ago         \\
    Probabilities          & 0.015 & 0.013 & 0.013 & 0.010 & 0.010 & 0.008   & 0.008 & 0.007 & 0.006       & 0.006 \\
    
    \bottomrule
    \end{tabular}
}
    \end{subtable}	
    \begin{subtable}[b]{\textwidth}
    	\centering
    	\caption{Music Piracy.}
    	\label{tab:2p2}
    	\resizebox{\textwidth}{10mm}{
    \begin{tabular}{BCr^r^r^r^r^r^r^r^r^r^r}
    \toprule
    \rowstyle{\bfseries}
      & w1    & w2    & w3    & w4    & w5    & w6      & w7    & w8    & w9          & w10  \\
    \midrule 
    Words                  & play  & music  & game  & nfl   & thank  & beat & best   & start  & soundcloud & anthem         \\
    Probabilities          & 0.016 & 0.012 & 0.012 & 0.011 & 0.011 & 0.009   & 0.009 & 0.009 & 0.007       & 0.007 \\
    
    \bottomrule
    \end{tabular}
}
    \end{subtable}
    \begin{subtable}[b]{\textwidth}
    	\centering
    	\caption{North Korean.}
    	\label{tab:2p3}
    	\resizebox{\textwidth}{10mm}{
    \begin{tabular}{BCr^r^r^r^r^r^r^r^r^r^r}
    \toprule
    \rowstyle{\bfseries}
      & w1    & w2    & w3    & w4    & w5    & w6      & w7    & w8    & w9          & w10  \\
    \midrule 
    Words                  & north  & korea  & anti  & mccain   & obamacare  & traitor & sessions   & fake  & attack & jeff         \\
    Probabilities          & 0.018 & 0.017 & 0.014 & 0.008 & 0.008 & 0.008   & 0.008 & 0.007 & 0.006       & 0.005 \\
    
    \bottomrule
    \end{tabular}
}
    \end{subtable}
    \begin{subtable}[b]{\textwidth}
    	\centering
    	\caption{Terrorism.}
    	\label{tab:2p4}
    	\resizebox{\textwidth}{10mm}{
    \begin{tabular}{BCr^r^r^r^r^r^r^r^r^r^r}
    \toprule
    \rowstyle{\bfseries}
      & w1    & w2    & w3    & w4    & w5    & w6      & w7    & w8    & w9          & w10  \\
    \midrule 
    Words                  & group  & terrorist  & expose  & terror   & honor  & fraud & islamic   & isis  & antifa & wikileaks         \\
    Probabilities          & 0.012 & 0.009 & 0.008 & 0.008 & 0.006 & 0.006   & 0.006 & 0.006 & 0.006       & 0.006 \\
    
    \bottomrule
    \end{tabular}
}
    \end{subtable}
    \caption{Useful topics.}
    \label{tab:2p}
\end{table*}
It is noticeable that the modified algorithm constructs many meaningful topics such as Supporting Hillary, Terrorism and so on, which means the performance of algorithm is improved. In general, the accuracy of the model is 35\%. On the other hand, we still find it difficult to recognize theme of some topics, for example table~\ref{tab:2n}. \\
\begin{table*}[h]
	\centering
	\caption{A useless topic (These words do not share any coherent theme).}
	\label{tab:2n}
	\resizebox{\textwidth}{10mm}{
    \begin{tabular}{BCr^r^r^r^r^r^r^r^r^r^r}
    \toprule
    \rowstyle{\bfseries}
      & w1    & w2    & w3    & w4    & w5    & w6      & w7    & w8    & w9          & w10  \\
    \midrule 
    Words                  & follow  & twitter  & wall  & thank   & understand  & matter & west   & quote  & build & party         \\
    Probabilities          & 0.018 & 0.013 & 0.011 & 0.011 & 0.010 & 0.009   & 0.008 & 0.007 & 0.007       & 0.006 \\
    
    \bottomrule
    \end{tabular}
}
\end{table*}\\
We change and test various parameter settings for the LDA algorithm, and the results are compared in the following table~\ref{tab:ttm}. Notice that decay is a number between 0.5 to 1.0 to weight what percentage of the previous lambda value is forgotten when each new document is examined \citep{hoffman2010online}.  Gamma\_threshold is the minimum change in the value of the gamma parameters to continue iterating. Passes is the number of passes through the corpus during training.\\
\begin{table*}[t!]
	\centering
	\caption{LDA model accuracy for different parameter settings.}
	
	\label{tab:ttm}
	% Every table you want to use the bold macros in must start with the B specifier.
% The C specifier makes the corresponding column bold. (so starting with "BCr"
% makes the first column bold and right-aligned)
% You must insert ^ between each column specifier.
% Then, the \rowstyle{\bfseries} turns that row bold.

\begin{tabular}{Br^r^r^r^r^r}
\toprule
\rowstyle{\bfseries}	
NumTopics	& Iterations  & Decay   	&	Gamma\_threshold	&	Passes	&	Accuracy\\
\midrule 
20      & 50        & 		0.5      	&		0.001		&		1		&		0.35\\
20      & 50        & 		0.6      	&		0.001		&		1		&		0.30\\
20      & 50        & 		0.7      	&		0.001		&		1		&		0.40\\
\midrule 
20      & 50        & 		0.7       	&		0.005		&		1		&		0.35\\
20      & 50        & 		0.7       	&		0.0005		&		1		&		0.50\\
20      & 50        & 		0.7       	&		0.0001		&		1		&		0.45\\
\midrule 
20      & 50        & 		0.7       	&		0.0005		&		10		&		0.40\\
20      & 50        & 		0.7       	&		0.0005		&		5		&		0.60\\
\midrule 
20      & 100        & 		0.7       	&		0.0005		&		5		&		0.60\\
20      & 200        & 		0.7       	&		0.0005		&		5		&		0.55\\
\midrule 
30      & 100        & 		0.7       	&		0.0005		&		5		&		0.53\\
10      & 100        & 		0.7       	&		0.0005		&		5		&		0.50\\

\bottomrule
\end{tabular}\\

%%% Local Variables: 
%%% mode: latex
%%% TeX-master: t
%%% End: 

\end{table*} \\
Again, the accuracy presented above are based on qualitative analysis of whether we can conclude a theme from a generated topic, and then we calculate the percentage of topics that can be concluded with a theme. There might be errors in the qualitative analysis, as the themes are concluded manually, thus, the accuracy are approximate values. From table~\ref{tab:ttm}, we know that the algorithm has the best performance with parameters that \icode{numTopic} is 20, \icode{iteration} is 100, \icode{decay} is 0.7, \icode{gamma\_threshold} is 0.0005, and \icode{passes} is 5. With the evaluation, we understand that the number of topic depends on the size of corpus and variety of documents. A large or low value will decrease the performance of the algorithm. We manually observed the content of these topics with that setting, and the extracted themes are: \textit{Supporting Trump, Against Hillary, Civil Rights, Music Piracy, Police shootings, School shootings, Islam and War, BlackLivesMatter, Terrorism, North Korea, Social Media Funneling,} and \textit{Protest Against Police}. In some topics, it is difficult to recognize the attitude within the theme, and in other words, we do not know whether trolls construct this topic in favour of Trump or Hillary. While the intention of trolls in some topics are clear, for example: table~\ref{tab:3d} can be concluded with a theme that \textit{Supporting Trump}, and table~\ref{tab:3h} contains the theme \textit{Against Hillary}. We need to evaluate other techniques that we implemented in order to further analyse the strategies of Russian troll.\\
\begin{table*}[t!]
	\centering
	\caption{Supporting Donald Trump.}
	\label{tab:3d}
	\resizebox{\textwidth}{10mm}{
    \begin{tabular}{BCr^r^r^r^r^r^r^r^r^r^r}
    \toprule
    \rowstyle{\bfseries}
      & w1    & w2    & w3    & w4    & w5    & w6      & w7    & w8    & w9          & w10  \\
    \midrule 
    Words                  & follow  & twitter  & wall  & thank   & understand  & matter & west   & quote  & build & party         \\
    Probabilities          & 0.018 & 0.013 & 0.011 & 0.011 & 0.010 & 0.009   & 0.008 & 0.007 & 0.007       & 0.006 \\
    
    \bottomrule
    \end{tabular}
}
\end{table*}\\
\begin{table*}[t!]
	\centering
	\caption{Against Hillary Clinton.}
	\label{tab:3h}
	\resizebox{\textwidth}{10mm}{
    \begin{tabular}{BCr^r^r^r^r^r^r^r^r^r^r}
    \toprule
    \rowstyle{\bfseries}
      & w1    & w2    & w3    & w4    & w5    & w6      & w7    & w8    & w9          & w10  \\
    \midrule 
    Words                  & believe  & nar  & yes  & muslims   & democrats  & life & neverhillary   & fail  & story & act         \\
    Probabilities          & 0.022 & 0.016 & 0.013 & 0.008 & 0.008 & 0.008 & 0.006   & 0.006 & 0.005 & 0.005        \\
    
    \bottomrule
    \end{tabular}
}
\end{table*}\\

\section{Supervised topic model}
\label{sec:estm}

\begin{figure*}[h]
	\centering
	\includegraphics[width=\textwidth*4/5]{{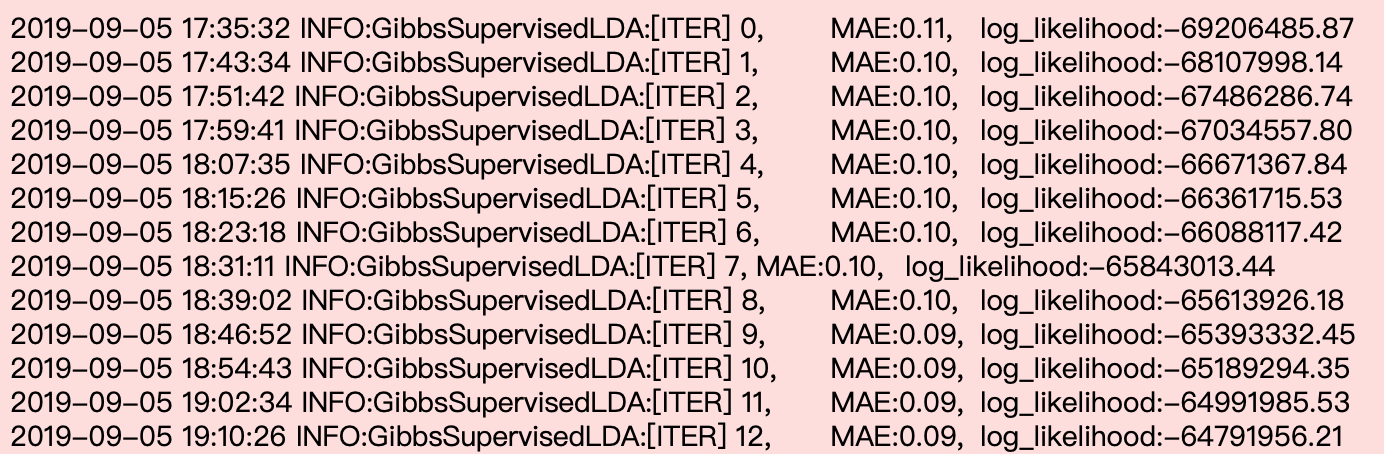}}
	\caption{Training log.}
	\label{fig:sldal}
\end{figure*}

\begin{figure*}[h]
	\centering
	\includegraphics[width=\textwidth]{{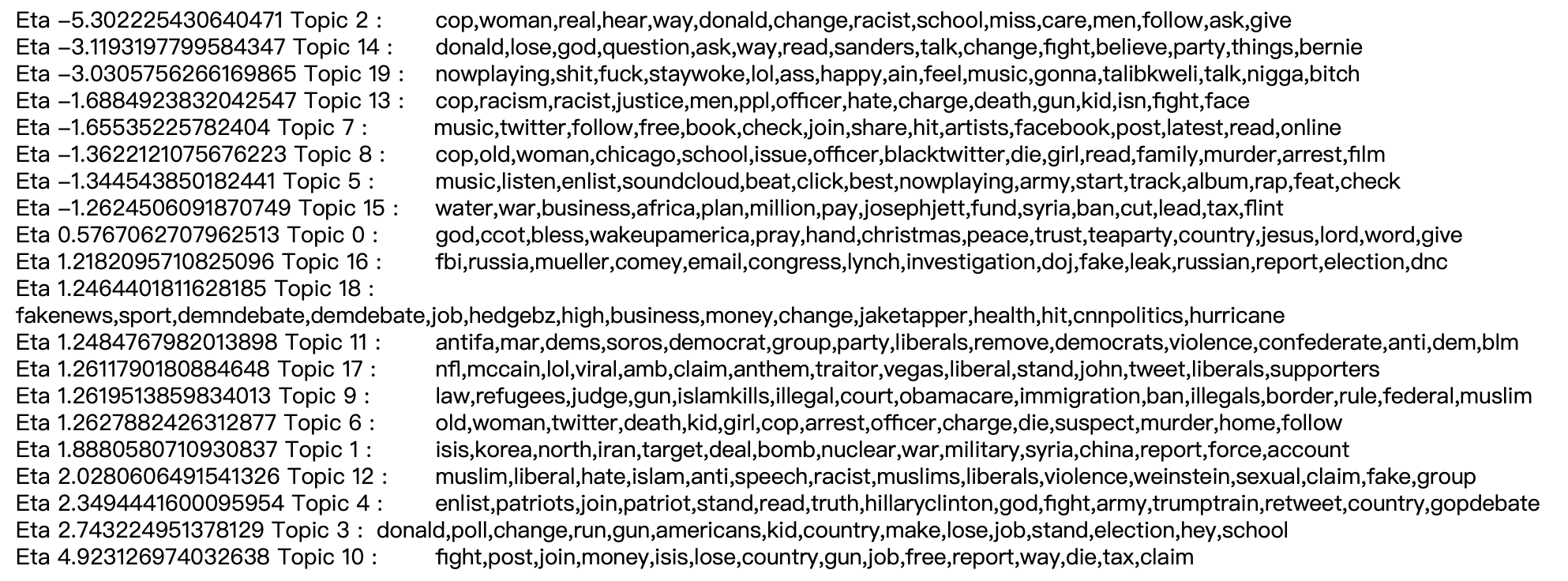}}
	\caption{Result of sLDA.}
	\label{fig:sldar}
\end{figure*}

In supervised topic model, we implement supervised Latent Dirichlet allocation (sLDA) with Gibbs sampling model. The Russian troll dataset is separated to training set and testing set, and we train the sLDA using training set, which predict troll category for each generated topic, and then predict labels for testing dataset. Mean  absolute error (mae) is calculated to evaluate the performance of sLDA by comparing predicted labels with ground true label provided in the test dataset. Additionally, we calculate the mae and log likelihood for each iteration when training the training set, and the number of iteration that we choose is a moderate number 100. We show part of learning records during the training in figure~\ref{fig:sldal}. It is apparently that the algorithm is continuously progressing, since the mae and negative log likelihood are decreasing, which means the algorithm is working and slowly learning patterns from the dataset. An example of predicted troll category with topics is figure~\ref{fig:sldar}\\
\\
\\
Eta is the predicted label that -1 represents left troll and 1 represents right troll, and the predicted value is not a binary category either -1 or 1. In fact, the model predicts a probability of being a left troll or right troll, which is a continuous number. In figure~\ref{fig:sldar}, some values are greater than 1 such as 4.923 for topic 10, and some values are smaller than -1 such as -5.302 for topic 2, because these topic have strong possibilities of inclination to left troll or right troll. For topics 2 with eta -5.302, it has strong probability to be a left troll which talks about Hillary Clinton and the Democratic Party. The topic contains words such as "women", "cop", and "racist", which might talks about issues like \textit{women's right, equal treatment, racism, police shooting, BlackLivesMatter}. These issues are in favour of the Democratic Party. However, we still notice word like "donald" appears in this topic, and we do not know the attitude of word "donald", thus, we need sentiment analysis to study the polarities. Another reason is that  the left troll intends to spread divisive  opinions to distract the Democratic Party as a strategy. For the Republican Party, topic 3 in figure~\ref{fig:sldar} has meaningful words such as "donald", "americans", "country", "job". This topic support Trump with his election slogan \textit{Make America Great Again}, and theme like \textit{Increasing Employment Rate}. As well as topic 12 that contains words like "muslim", "islam", "anti", and these words represent themes that \textit{Anti-Muslim, Immigration Policy}, which are strategies for the Republican Party. \\\\
Then we calculate the mae to measure the performance of our sLDA algorithm on the testing dataset. The mae will be different even with the same parameter and number of topic, thus we record and calculate the average mae for running ten times. We change sigma in the prediction update algorithm in the sLDA model, compared in table~\ref{tab:slda}. Therefore, we choose setting No. 3 with sigma 0.01. The predicted troll labels have higher probabilities for some topics, table~\ref{tab:sh} illustrates topics with high possibility to be left troll broadcasting themes such as \textit{women's right, equal treatment, racism, police shooting, BlackLivesMatter, music piracy}. Table~\ref{tab:sd} illustrates topics with high possibility to be right troll broadcasting themes such as \textit{Make America Great Again, Increasing Employment Rate, War/Military, Islam, Terrorism}, and we also found that \textit{Crimes} theme focuses on crimes against children and women.\\
\begin{table*}[h]
	\centering
	\caption{sLDA model mae for different parameter settings.}
	\label{tab:slda}
	% Every table you want to use the bold macros in must start with the B specifier.
% The C specifier makes the corresponding column bold. (so starting with "BCr"
% makes the first column bold and right-aligned)
% You must insert ^ between each column specifier.
% Then, the \rowstyle{\bfseries} turns that row bold.

\begin{tabular}{BCr^r^r}
\toprule
\rowstyle{\bfseries}
 No.      & Sigma  & MAE average\\
\midrule 
1              & 	0.1	& 0.9945     \\
2              & 	0.05	& 0.9804     \\
3              & 	0.01	& 0.9754     \\
4              & 	0.005	& 0.9832     \\

\bottomrule
\end{tabular}

%%% Local Variables: 
%%% mode: latex
%%% TeX-master: t
%%% End: 

\end{table*} \\
\begin{table*}[h]
	\centering
	\caption{Topics strongly related to the left trolls.}
	\label{tab:sh}
	\resizebox{\textwidth}{8mm}{
    \begin{tabular}{Br^r^r^r^r^r^r^r^r^r^r}
    \toprule
    \rowstyle{\bfseries}
       Eta  & Topic      & Top probability words    \\
    \midrule 
    -7.97                  & 1  & start, womam, best, cop, way, real, school, donald, free, change, racist, things, remember, make, listen \\
    -5.65                  & 5  & womam, start, cop, hear, change, way, better, men, real, follow, best, free, music, join, read \\
    
    \bottomrule
    \end{tabular}
}
\end{table*} 
\begin{table*}[h]
	\centering
	\caption{Topics strongly related to the right trolls.}
	\label{tab:sd}
	\resizebox{\textwidth}{8mm}{
    \begin{tabular}{Br^r^r^r^r^r^r^r^r^r^r}
    \toprule
    \rowstyle{\bfseries}
       Eta  & Topic      & Top probability words    \\
    \midrule 
    3.63                  & 10  & home, local, old, die, dead, feel, god, make, woman, texas, little, dog, christmas, kid, away\\
    -6.01                  & 11  & nfl, country, job, donald, national, gun, liberals, fake, change, korea, lose, liberal, claim, isis, anthem\\
    
    \bottomrule
    \end{tabular}
}
\end{table*}\\
We implement restricted time range in the dataset, and train the sLDA model separately. We compared the mae of sLDA model with dataset in 2015, 2016, and 2017 in table~\ref{tab:sldat}. Low mae value means that prediction of troll category for topics are more precise. The accuracy changes when the time range is restricted with the same parameter settings. That might because when the training dataset is more sparse, which means contains more themes, and topics change over time, the model is more difficult to learn from the dataset, and thus, performs bad on prediction. The is the reason why the performances of models using restricted time range are generally better than original model using the whole dataset. Therefore, we conclude that the strategies of Russian trolls change during 2015 to 2016.
\begin{table*}[h]
	\centering
	\caption{sLDA model mae with different time range.}
	\label{tab:sldat}
	% Every table you want to use the bold macros in must start with the B specifier.
% The C specifier makes the corresponding column bold. (so starting with "BCr"
% makes the first column bold and right-aligned)
% You must insert ^ between each column specifier.
% Then, the \rowstyle{\bfseries} turns that row bold.

\begin{tabular}{BCr^r^r}
\toprule
\rowstyle{\bfseries}
 No.      & Year  & MAE average\\
\midrule 
1              & 2015        & 0.8341     \\
2              & 2016        & 0.9831     \\
3              & 2017        & 0.7415     \\

\bottomrule
\end{tabular}

%%% Local Variables: 
%%% mode: latex
%%% TeX-master: t
%%% End: 

\end{table*} \\

\section{Sentiment analysis}
\label{sec:esa}

\begin{figure*}[h]
	\centering
	\includegraphics[width=\textwidth*4/5]{{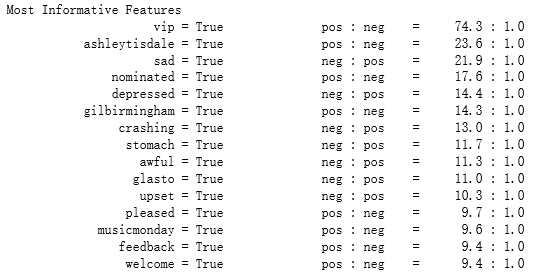}}
	\caption{Informative features in sentiment classifier}
	\label{fig:1}
\end{figure*}

We implement sentiment analysis to observe attitudes of tweets in interesting topics. A variety of classification models are used in sentiment classification, and the Naive Bayes classifier is the baseline model. As we mentioned before, we train the model using an Twitter sentiment dataset rather than the Russian troll dataset, and the dataset we used is extremely large containing 1.6 million tweets. Therefore, we only extract part of the data to train the classifier. Initially, we extract 50,000 tweets from both negative and positive tweets datasets, and we pre-process the dataset and train the model using Naive Bayes Classifier. The result is presented in figure~\ref{fig:1}. We list several most informative features such as "vip", "nominated", "pleased", "welcome" are positive words and "sad", "awful", "crashing" are negative. Then we compared different classifiers by evaluating classification accuracy and F1$_{micro}$ score, and the results are presented in table~\ref{tab:sa}. \\
\\
\begin{table*}[h]
	\centering
	\caption{Sentiment classifiers performances for different parameter settings.}
	\label{tab:sa}
	% Every table you want to use the bold macros in must start with the B specifier.
% The C specifier makes the corresponding column bold. (so starting with "BCr"
% makes the first column bold and right-aligned)
% You must insert ^ between each column specifier.
% Then, the \rowstyle{\bfseries} turns that row bold.

\begin{tabular}{BCr^r^r}
\toprule
\rowstyle{\bfseries}
  Classifier     & Accuracy  percent  & F1 score  \\
\midrule 
Naive Bayes              & 70.26        & 0.7026         \\
Multinomial Naive Bayes  & 71.01 &0.7101 \\
Bernoulli Naive Bayes  & 71.045 &0.71045 \\
Logistic Regression & 71.67 &0.7167\\
Stochastic Gradient Descent & 70.92 &0.7092\\
\bottomrule
\end{tabular}

%%% Local Variables: 
%%% mode: latex
%%% TeX-master: t
%%% End: 

\end{table*} \\
The sentiment model produces a binary category that either positive or negative. We ensemble these classifiers to one model, and count the votes from different classifiers as confidence. Thus, a high confidence means it is more possible to be that category, as the majority of classifiers predict that category. Note that the confidence ranges from 0.5 to 1.0, and because there are only 5 classifier, thus, the confidence will only be 0.6, 0.8, or 1.0.\\\\
After constructing an efficient sentiment classifier, we predict sentiment on representative tweets extracted from interesting topics with high possibility to be left troll or right troll. For example, in figure~\ref{fig:sldar}, topic 10 has eta 4.923, which means topic 10 is more likely to be right troll topic supporting Trump, and attacking Hillary Clinton. Tweet in table~\ref{tab:eg1} contains a theme attacking Hillary Clinton, and the sentiment is classified to negative attitude with 1.0 confidence. Tweets in table~\ref{tab:eg2} and  table~\ref{tab:eg3} generate positive sentiment prediction that supporting Trump with high confidence, which correspond to the meaning of tweets. Therefore, we recognize that right troll supports Donald Trump and against Hillary Clinton.\\
\begin{table*}[h]
    \begin{subtable}[b]{\textwidth}
    	\centering
    	\caption{Against Hillary Clinton.}
    	\label{tab:eg1}
    	
\resizebox{\textwidth}{9mm}{
    
    \begin{tabular}{Br^r^r^r^r^r^r^r^r^r^r}
    
    \toprule
    \rowstyle{\bfseries}
      & Content    & Sentiment label    & Confidence    \\
    \midrule 
     "It took Hillary abt 5 minutes to blame NRA for madman's rampage, & \multirow{3}[2]{*}{'neg'}   & \multirow{3}[2]{*}{1.0} \\
     but 5 days to sorta-kinda blame Harvey Weinstein 4 his sextually assaults."   &  & \\
    
    \bottomrule
    \end{tabular}
}
    \end{subtable}	
    
    \vspace{1em}
    
    \begin{subtable}[b]{\textwidth}
    	\centering
    	\caption{Supporting Trump.}
    	\label{tab:eg2}
    	
\resizebox{\textwidth}{9mm}{
    
    \begin{tabular}{Br^r^r^r^r^r^r^r^r^r^r}
    
    \toprule
    \rowstyle{\bfseries}
      & Content    & Sentiment label    & Confidence    \\
    \midrule 
     "Trump wants to give low income balck parents the freedom to choose  & \multirow{3}[2]{*}{'pos'}   & \multirow{3}[2]{*}{1.0} \\
     where their kid go to school. Democrats don't."   &  & \\
    
    \bottomrule
    \end{tabular}
}
    \end{subtable}

    \vspace{1em}
        
    \begin{subtable}[b]{\textwidth}
    	\centering
    	\caption{Supporting Trump.}
    	\label{tab:eg3}
    	
\resizebox{\textwidth}{9mm}{
    
    \begin{tabular}{Br^r^r^r^r^r^r^r^r^r^r}
    
    \toprule
    \rowstyle{\bfseries}
      & Content    & Sentiment label    & Confidence    \\
    \midrule 
     "THIS IS AWESOME! While Hillary's in her BUNKER plugged full of IVs, & \multirow{3}[2]{*}{'pos'}   & \multirow{3}[2]{*}{1.0} \\
     Trump is talking the ECONOMY. Wait til end."   &  & \\
    
    \bottomrule
    \end{tabular}
}
    \end{subtable}
    \caption{Sentiment analysis for topic 10 (Right troll).}
    \label{tab:saeg1}
\end{table*}\\ 
Table~\ref{tab:saeg2} represents tweets extracted from topic 2 broadcasted by right troll. Tweet in table~\ref{tab:egh} contains a theme supporting Hillary Clinton, which corresponds to the predicted sentiment. The sentiment analysis makes mistakes in some situations. Tweet in table~\ref{tab:egdn} contains a theme against Trump, but the predicted sentiment of this tweet is positive. The reason might be that the tweet uses sarcasm, and the attitude is latent, since there are no negative words appear in the tweet. Therefore, the sentiment classifier cannot distinguish polarity from the content.
\begin{table*}[h]
    \begin{subtable}[b]{\textwidth}
    	\centering
    	\caption{Supporting Hillary Clinton.}
    	\label{tab:egh}
    	
\resizebox{\textwidth}{8mm}{
    
    \begin{tabular}{Br^r^r^r^r^r^r^r^r^r^r}
    
    \toprule
    \rowstyle{\bfseries}
      & Content    & Sentiment label    & Confidence    \\
    \midrule 
     "I believed in her then, I believe in her now. HillaryClinton  &'pos'   &1.0 \\
    
    \bottomrule
    \end{tabular}
}
    \end{subtable}	

    \vspace{1em}
        
    \begin{subtable}[b]{\textwidth}
    	\centering
    	\caption{Against Trump.}
    	\label{tab:egdn}
    	
\resizebox{\textwidth}{9mm}{
    
    \begin{tabular}{Br^r^r^r^r^r^r^r^r^r^r}
    
    \toprule
    \rowstyle{\bfseries}
      & Content    & Sentiment label    & Confidence    \\
    \midrule 
     "Incredible courge. Now 18 year olds need puppies and coloring books  & \multirow{3}[2]{*}{'pos'}   & \multirow{3}[2]{*}{0.8} \\
     on campus because Trump was elected. Sofa King nut."   &  & \\
    
    \bottomrule
    \end{tabular}
}
    \end{subtable}
    \caption{Sentiment analysis for topic 2 (Left troll).}
    \label{tab:saeg2}
\end{table*}

\section{Summary}
We evaluate our implemented models and illustrate the results in this chapter. We have found some interesting topics contain several themes using LDA model. We find strategies of right troll and left troll from generated topic with predicted troll category, and we also observe that the strategies of Russian trolls change over time from sLDA model. We further understand troll's strategies in each topic by observing attitudes of tweets using  sentiment analysis. In next chapter, we will discuss the results we discovered, and answer the questions we proposed in section\ref{sec:motivations}.
\chapter{Discussion}
\label{cha:discussion}
In this chapter, we interpret the results generated from models that we implement, and answer the two questions that: 
\begin{enumerate}
	\item How to understand the behaviour of different types of Russian trolls?
	\item How to understand the behaviour and strategy of Russian trolls change over time?
\end{enumerate}
For the first question, we present initial analyses of behavioural patterns, and thematic content retrieved from the from the Russian troll dataset, containing Twitter content propagated by the Internet Research Agency (IRA) on behalf of Russian political interests. We observe themes from LDA and sLDA models such as \textit{Civil Rights, Police shootings, Music Piracy, Islam and War, Military, Supporting Trump, Against Trump, BlackLivesMatter, Women’s Rights, Social Media Funneling, Supporting Hillary, Against Hillary, Gun Control }and \textit{Crimes}. Our analyses imply that the 2016 U.S. Presidential Election was influenced by IRA controlled by Russian government. They manipulate Russian social media trolls to attempt to dominate public opinions in these hot topics that we extracted. Our generated themes are similar with results concluded from the Facebook advertised posts in \citep{boyd2018characterizing}, which emphasizes that the IRA appears to use similar patterns in both social media platforms.\\\\
We notice that trolls are work together. Topics that left troll and right troll focused overlap a lot, and their opinions various in each topic. For example, in some themes, like \textit{Supporting Trump} and \textit{Attacking Hillary}, contradicted opinions are found in favor and against the main themes from sentiment analysis. Trolls try to spread polarizing content. Moreover, in some situations, we find both right trolls and left trolls have similar or related topics from our sLDA model, even have the same attitudes.\\\\
Another finding is that left trolls have a more complex discursive strategy, while right trolls have a relatively more homogeneous identity. Russian trolls have more attention to the ideological right. Although there are fewer number of topics generated for left troll than from the right by sLDA, we observe more themes from left troll topics rather than right troll. In some areas left trolls support Democratic through BlackLivesMatter, women’s rights and other debates, in others they attack Hillary and talk about issues such as war, religious identity, and crimes. Although topics of left troll and right troll intersect a lot, right trolls relatively focus more on mainstream Republicanism. They have themes of \textit{supporting Trump, gun control, war, Muslims} which all generally in favor of ideological right. The number of predicted topics produced by right troll is generally greater than topics from left troll. Therefore, we conclude that left troll is more divisive than right troll, since the IRA intends to support the ideological right.\\\\For the second question, the analysis of thematic content with restricted time range in the Russian dataset allows us to better understand the strategies of trolls change from 2015 to 2017. We observe the mean absolute error (mae) of sLDA with dataset containing tweets published in 2016 is higher than mae in 2017, 2015, and whole dataset. The election was occurred in 2016, and the IRA try to spread divisive opinions to manipulate the election. More specifically, they particularly focus on spreading a high amount of polarizing content by both left troll and right troll to frustrate the ideological left. Evidence is shown that the number of generated topics are evenly distributed among left troll and right troll. The mae of sLDA model for dataset in 2017 is much lower than that in 2016, that might because they change strategies to support the mainstream Republicanism. In sLDA result for 2017, we observe that right troll generates the majority of topics, and the themes extracted from topics are more homogeneous focusing on the ideological right with the help of sentiment analysis.

\chapter{Conclusion}
\label{cha:conc}

In this work, we analyse the strategies that Russian trolls used during 2016 U.S. Presidential Election from a publicly available Twitter dataset containing tweets published by Russian trolls. The aim is finding the evidence that whether Russian trolls interfered with the election, and how different types of troll work, and how are the strategies change over time. A variety of natural language processing techniques are implemented. Firstly, Latent Dirichlet allocation (LDA) is applied for analysing topic model of the dataset. And then we implement supervised Latent Dirichlet allocation (sLDA) to predict types of troll categories for each generated topic. Additionally, we restrict the time range of tweets in the dataset, followed by applying sLDA to further analyse how trolls change their behaviours over time. Finally, we do sentiment analysis to predict the polarity of tweets from interested topics. The attitudes of troll in each predicted topic offer a holistic view of strategies of Russian trolls. Furthermore, we observe that Russian trolls work together, and they generally have pro-Trump, conservative political agenda during 2016 U.S. election. The right trolls behave like typical Trump supporters by broadcasting news that benefit his election. While the left troll act derisively towards Hillary Clinton by broadcasting divisive content in the Democracy Party. In 2016, Russian trolls have a strategy of attempting to spread a large amount of divisive opinions to intervene the election. After the 2016 election, they change their strategies to focus on supporting the ideological right by broadcasting pro-Trump materials with a relatively more homogeneous identity.

\section{Future Work}
\label{sec:future}
This research has limitations which can be addressed in the future:
\begin{enumerate}
    \item We only analyse strategies of Russian trolls in terms of right troll and left troll from Russian troll dataset in this paper. However, there exist other types of trolls such as news feed, and hashtag gamer. Studies for other troll categories will conducted in the future.
    
    \item In this research, we use a qualitative way to conclude themes from topics based on human intuition, and in the future, we will propose an automated way to distinguish meaningful topics by using machine learning methods.

\end{enumerate}

%%%%%%%%%%%%%%%%%%%%%%%%%%%%%%%%%%%%%%%%%%%%%%%%%%%%%%%%%%%%%%%%%%%%%%
% Here begins the end matter

\backmatter

%%%%%%%%%%%%%%%%%%%%%%%%%%%%%%%%%%%%%%%%%%%%%%%%%%%%%%%%%%%%%%%%%%%%%%
%% Bibliography
\cleardoublepage
\phantomsection
\addcontentsline{toc}{chapter}{Bibliography}
\bibliographystyle{anuthesis}
\bibliography{report}

\printindex

\end{document}